\documentclass[12pt]{article}
\usepackage{epsfig}

\date{}
\begin{document}

\newcommand{\beq}{\begin{equation}}
\newcommand{\eeq}{\end{equation}}
\newcommand{\nn}{\nonumber}
\newcommand{\bea}{\begin{eqnarray}}
\newcommand{\eea}{\end{eqnarray}}

\title{Modified Gravity on the Brane and Dark Energy}

\author{Emilio Elizalde\footnote{E-mail:
  elizalde@ieec.fcr.es}\\{\small
  \it Instituto de Ciencias del Espacio (CSIC) }\\
{\small \it \&
Institut d'Estudis Espacials de Catalunya (IEEC/CSIC) }\\
{\small \it
Campus UAB, Facultat de Ci\`encies, Torre C5-Parell-2a planta }\\
{\small \it
E-08193 Bellaterra (Barcelona), Spain}\\
\mbox{}\\
Rui Neves\footnote{E-mail: rneves@ualg.pt. Also at Centro Multidisciplinar de Astrof\'{\i}sica - CENTRA, Instituto Superior
  T\'ecnico, Avenida Rovisco Pais, 1049-001 Lisboa and Departamento de F\'{\i}sica, Faculdade de Ci\^encias e Tecnologia,
  Universidade do Algarve, Campus de Gambelas, 8005-139 Faro, Portugal.}\\
{\small \it Centro de Electr\'onica, Optoelectr\'onica e Telecomunica\c {c}\~oes (CEOT)}\\
{\small \it Faculdade de Ci\^encias e Tecnologia}\\
{\small \it  Universidade do Algarve}\\
{\small \it Campus de Gambelas, 8005-139 Faro, Portugal}
}

\maketitle

\begin{abstract}

We analyze the dynamics of an $\mbox{AdS}_5$ braneworld with matter
fields when gravity is allowed to deviate from the Einstein form on
the brane. We consider exact 5-dimensional warped solutions which
are associated with conformal bulk fields of weight -4 and describe
on the brane the following three dynamics: those of inhomogeneous
dust, of generalized dark radiation, and of homogeneous polytropic
dark energy. We show that, with modified gravity on the brane, the
existence of such dynamical geometries requires the presence of
non-conformal matter fields confined to the brane.

\end{abstract}

\section{Introduction}

According to the Randall-Sundrum (RS) braneworld scenario
\cite{RS1,RS2} the visible Universe is a 3-brane boundary of a $Z_2$
symmetric 5-dimensional (5D) anti-de Sitter (AdS) space. In the RS1
model \cite{RS1} the $\mbox{AdS}_5$ orbifold has a compact fifth
dimension and two branes. Gravity is localized on the hidden
positive tension brane and decays towards the visible negative
tension brane. In this configuration the RS scenario allows the
hierarchy problem to be reformulated as an exponential hierarchy
between the weak and Planck scales \cite{RS1}. In the RS2 model
\cite{RS2} the fifth dimension is infinite and there is just one
visible positive tension brane to which gravity is exponentially
bound.

On the visible brane, the low energy theory of gravity is 4D
Einstein general relativity and the cosmology may be
Friedmann-Robertson-Walker \cite{RS1}-\cite{TM}. With two branes
this requires the stabilization of the radion mode by introducing,
for example, a bulk scalar field \cite{GW,WFGK,CGRT,TM}. Using the
Gauss-Codazzi formulation \cite{BC,SMS} several other braneworld
solutions have been discovered although a number of them have not
yet been associated with exact 5D spacetimes \cite{COSp}-\cite{RC1}.

In this paper we continue the analysis of the dynamics of a
spherically symmetric RS 3-brane when conformal matter fields
propagate in the bulk \cite{RC2}-\cite{RC4} (see also \cite{EONO}).
Some time ago, a new class of exact 5D dynamical warped solutions
was discovered which is associated with conformal fields
characterized by an energy-momentum tensor of weight -4. These
solutions were shown to describe on the brane the dynamics of
inhomogeneous dust, generalized dark radiation and homogeneous
polytropic matter \cite{RC2,RC3}. In addition, the radion mode was
stabilized by a sector of the conformal bulk fluid with a constant
negative 5D pressure \cite{RN,RC4}. In this work we explore the
possibility of modifying the theory of gravity on the brane \cite{NO1,modg}
and we analyze if, after that change, the model is still able to
generate dynamics of these kinds, in particular the relevant one of
a perfect dark energy fluid describing the accelerated expansion of
our Universe. As we shall see, the answer will be in the
affirmative, under some conditions that will be made explicit in
what follows.

\section{5D equations with modified gravity and\\ conformal fields}

In a 5D RS orbifold mapped by a set of comoving coordinates,
$(t,r,\theta,\phi,z)$, the most general dynamical metric consistent
with the $Z_2$ symmetry in $z$ and with 4D spherical symmetry on the
brane is given by \beq
d{\tilde{s}_5^2}={\Omega^2}\left(d{z^2}-{e^{2A}}d{t^2}+{e^{2B}}d{r^2}
+{R^2}d{\Omega_2^2}\right),\label{eq1} \eeq where $A=A(t,r,z)$,
$B=B(t,r,z)$, $R=R(t,r,z)$ and $\Omega=\Omega(t,r,z)$ are $Z_2$
symmetric functions. $R$ represents the physical radius of the
2-spheres, $\Omega$ is the warp factor which defines a global
conformal transformation on the 5D metric, and
$d{\Omega_2^2}=d{\theta^2}+{\sin^2}\theta d{\phi^2}$.

Let us consider a dynamical RS action with matter fields and allow
gravity on the brane to differ from the standard Einstein-Hilbert
form. In the RS2 setting, we write
\[
\tilde{\mathcal{S}}=\int{d^4}xdz\sqrt{-\tilde{g}}
\left\{{\tilde{R}\over{2{\kappa_5^2}}}-{\Lambda_{\rm
B}}+\left[-{\lambda\over{\sqrt{\tilde{g}_{55}}}}
+{{\tilde{\mathcal{K}}}\over{\kappa_5^2}}
+f(\tilde{R}_4)\right]\delta(z-{z_0})\right\}
\]
\beq
+\int{d^4}xdz\sqrt{-\tilde{g}} {\tilde{\mathcal{L}}_{\rm
M}},\label{5Dact}
\eeq
where $\Lambda_{\rm B}$ is the
negative bulk cosmological constant, ${\kappa_5^2}=8\pi/{M_5^3}$
with $M_5$ the fundamental 5D Planck mass, and $f$ is some arbitrary
function of the 4D Ricci scalar, $\tilde{R}_4$. A special example is the
modified gravity theory with
\beq
f(\tilde{R}_4)=-{a\over{\tilde{R}_4}}+b{\tilde{R}_4^2},\label{spf}
\eeq
which as been considered as a gravitational alternative for the
cosmic dark energy fluid \cite{NO1}. The brane is located
at $z={z_0}$ and has tension $\lambda$. Note that $\lambda$ is in
fact the zero mode of $f$. The contribution of the matter fields is
defined by the lagrangian $\tilde{\mathcal{L}}_{\rm M}$. These may
include sectors which are confined to the brane,
\beq
{\tilde{\mathcal{L}}_{\rm M}}={\tilde{\mathcal{L}}_{\rm
B}}+{{\tilde{\mathcal{L}}_{\rm b}}\over{\sqrt{\tilde{g}_{55}}}} \,
\delta\left(z-{z_0}\right).
\eeq
In the Hawking-Gibbons term the
boundary scalar curvature is
$\tilde{\mathcal{K}}={\tilde{g}^{\mu\nu}}{\tilde{\mathcal{K}}_{\mu\nu}}$
where the extrinsic curvature is defined by
${\tilde{\mathcal{K}}_{\mu\nu}}={\tilde{\nabla}_\mu}{\tilde{n}_\nu}$.
The vector normal to the brane boundary is
${\tilde{n}_\nu}={\delta_\nu^5}\sqrt{\tilde{g}_{55}}$. In the RS1
model, the brane with tension $\lambda$ and gravity $f$ is the
hidden Planck brane. To represent the visible brane we introduce
another Dirac delta source at $z={{z'}_0}$ with a tension $\lambda'$
and a theory of gravity defined by a new function
$g({\tilde{R}_4})$. The analysis that follows is valid for both RS
models. For simplicity, we omit explicit reference to the second
brane in the RS1 setting.

A Noether variation on the action (\ref{5Dact}) gives the 5D
classical field equations
\[
{\tilde{G}_\mu^\nu}=-{\kappa_5^2}\left[{\Lambda_{\rm
B}}{\delta_\mu^\nu}+\delta\left(z-{z_0}\right)
\left(\lambda-{{\tilde{\mathcal{K}}}\over{\kappa_5^2}}
-f(\tilde{R}_4)\right){\tilde{\gamma}_\mu^\nu}\right]
\]
\beq
-{\kappa_5^2}\left[\delta\left(z-{z_0}\right){2\over{\sqrt{\tilde{g}_{55}}}}\left({{\tilde{\mathcal{K}}_\mu^\nu}\over{\kappa_5^2}}
+f'(\tilde{R}_4){\tilde{R}_{4\mu}^\nu}\right)-{\tilde{\mathcal{T}}_{{\rm
    M}\mu}^\nu}\right],
\label{5DEeq}
\eeq
where the prime denotes differentiation with
respect to the argument $x$ of the function $f(x)$. The induced
metric on the brane is
\beq
{\tilde{\gamma}_\mu^\nu}={1\over{\sqrt{\tilde{g}_{55}}}}\left(
{\delta_\mu^\nu}-{\delta_\mu^5}{\delta_5^\nu}\right)
\eeq
and the
stress-energy tensor $\tilde{\mathcal{T}}_{{\rm M}\mu}^\nu$
associated with the matter fields is defined by
\beq
{\tilde{\mathcal{T}}_{{\rm M}\mu}^\nu}={\tilde{\mathcal{L}}_{\rm
M}}{\delta_\mu^\nu}- 2{{\delta{\tilde{\mathcal{L}}_{\rm
M}}}\over{\delta{\tilde{g}^{\mu\alpha}}}}
{\tilde{g}^{\alpha\nu}},\quad{\tilde{\mathcal{T}}_{{\rm
      M}\mu}^\nu}={\tilde{\mathcal{T}}_{{\rm
      B}\mu}^\nu}+{{\tilde{\mathcal{T}}_{{\rm b}\mu}^\nu}
      \over{\sqrt{\tilde{g}_{55}}}}\delta\left(z-{z_0}\right).
\eeq
The bulk stress-energy tensor is conserved,
\beq
{\tilde{\nabla}_\mu}{\tilde{\mathcal{T}}_{{\rm
B}\nu}^\mu}=0.\label{5Dceq}
\eeq
Note that, in general,
${\tilde{\mathcal{T}}_{{\rm b}\mu}^\nu}$ is not conserved.

For a general 5D metric, $\tilde{g}_{\mu\nu}$, (\ref{5DEeq}) and
(\ref{5Dceq}) is difficult to solve system of partial differential
equations. Let us introduce some simplifying assumptions on the
field variables involved in the problem. First, consider that the
bulk matter is described by conformal fields with weight $s$. Under
the conformal transformation
${\tilde{g}_{\mu\nu}}={\Omega^2}{g_{\mu\nu}}$, the stress-energy
tensor satisfies ${\tilde{\mathcal{T}}_{{\rm
      B}\mu}^\nu}={\Omega^{s+2}}{\mathcal{T}_{{\rm B}\mu}^\nu}$.
Consequently, (\ref{5DEeq}) and (\ref{5Dceq}) may be re-written as
\cite{RC2}
\[
{G_\mu^\nu}=-6{\Omega^{-2}}\left({\nabla_\mu}\Omega\right){g^{\nu\rho}}
{\nabla_\rho}\Omega+
3{\Omega^{-1}}{g^{\nu\rho}}{\nabla_\rho}{\nabla_\mu}\Omega
-3{\Omega^{-1}}{\delta_\mu^\nu}{g^{\rho\sigma}}{\nabla_\rho}{\nabla_\sigma}
\Omega
\]
\[
-{\kappa_5^2}
{\Omega^2}{\Lambda_{\rm B}}{\delta_\mu^\nu}+{\kappa_5^2}{\Omega^{s+4}}{\mathcal{T}_{{\rm B}\mu}^\nu}-{\kappa_5^2} {\Omega^2}\delta \left(z-{z_0}\right)
\left(\lambda-{{\tilde{\mathcal{K}}}\over{\kappa_5^2}}
-f(\tilde{R}_4)\right){\tilde{\gamma}_\mu^\nu}
\]
\beq
-{\kappa_5^2} {\Omega^2}\delta \left(z-{z_0}\right){2\over{\sqrt{\tilde{g}_{55}}}}\left({{\tilde{\mathcal{K}}_\mu^\nu}
\over{\kappa_5^2}}+f'(\tilde{R}_4){\tilde{R}_{4\mu}^\nu}-{{\tilde{\mathcal{T}}_{{\rm
          b}\mu}^\nu}\over{2}}\right),
\label{t5DEeq}
\eeq
\beq
{\nabla_\mu}{\mathcal{T}_{{\rm
      B}\nu}^\mu}+{\Omega^{-1}}\left[(s+7){\mathcal{T}_{{\rm B}\nu}^\mu}{\partial_\mu}
\Omega-{\mathcal{T}_{{\rm B}\mu}^\mu}{\partial_\nu}\Omega\right]=0\label{t5Dceq}.
\eeq

If we split the bulk tensor $\tilde{\mathcal{T}}_{{\rm B}\mu}^\nu$
into two sectors, $\tilde{T}_{{\rm B}\mu}^\nu$ and $\tilde{U}_{{\rm
B}\mu}^\nu$, with the same weight $s$, ${\tilde{\mathcal{T}}_{{\rm
B}\mu}^\nu}={\tilde{T}_{{\rm
      B}\mu}^\nu}+{\tilde{U}_{{\rm B}\mu}^\nu}$,
where ${\tilde{T}_{{\rm B}\mu}^\nu}={\Omega^{s+2}}{T_{{\rm
B}\mu}^\nu}$ and ${\tilde{U}_{{\rm
B}\mu}^\nu}={\Omega^{s+2}}{U_{{\rm B}\mu}^\nu}$, and take $s=-4$,
then it is possible to separate (\ref{t5DEeq}) in the following way
\cite{RC4}
\beq
{G_\mu^\nu}={\kappa_5^2}{T_{{\rm
B}\mu}^\nu},\label{r5DEeq}
\eeq
\[
6{\Omega^{-2}}{\nabla_\mu}\Omega{\nabla_\rho}
\Omega{g^{\rho\nu}}-
3{\Omega^{-1}}{\nabla_\mu}{\nabla_\rho}\Omega{g^{\rho\nu}}+3{\Omega^{-1}}
{\nabla_\rho}{\nabla_\sigma}\Omega{g^{\rho\sigma}}{\delta_\mu^\nu}=
\]
\[
-{\kappa_5^2}\left[
{\Omega^2}{\Lambda_{\rm B}}{\delta_\mu^\nu}-
{U_{{\rm B}\mu}^\nu}+
{\Omega^2}\delta
\left(z-{z_0}\right)
\left(\lambda-{{\tilde{\mathcal{K}}}\over{\kappa_5^2}}
-f(\tilde{R}_4)\right){\tilde{\gamma}_\mu^\nu}\right]
\]
\beq
-{\kappa_5^2}
{\Omega^2}\delta
\left(z-{z_0}\right){2\over{\sqrt{\tilde{g}_{55}}}}\left({{\tilde{\mathcal{K}}_\mu^\nu}
\over{\kappa_5^2}}+f'(\tilde{R}_4){\tilde{R}_{4\mu}^\nu}-{{\tilde{\mathcal{T}}_{{\rm
          b}\mu}^\nu}\over{2}}\right).\label{5DEeqwf}
\eeq Within this splitting, the Bianchi identity implies \beq
{\nabla_\mu}{T_{{\rm B}\nu}^\mu}=0.\label{5DceqT} \eeq Then,
(\ref{t5Dceq}) is in fact \beq {\nabla_\mu}{U_{{\rm
      B}\nu}^\mu}+{\Omega^{-1}}\left(3{\mathcal{T}_{{\rm B}\nu}^\mu}{\partial_\mu}
\Omega-{\mathcal{T}_{{\rm B}\mu}^\mu}{\partial_\nu}\Omega\right)=0.\label{u5Dceq}
\eeq
In (\ref{r5DEeq}) and (\ref{5DceqT}) we have 5D
equations with conformal bulk fields but without
a brane or bulk cosmological constant. These equations do not depend on the
warp factor which is dynamically defined by (\ref{5DEeqwf}) and
(\ref{u5Dceq}). Consequently, in this simplified setting the warp
is the only effect reflecting the
existence of the brane or of the bulk cosmological constant. Note that
this separation is only possible with the special set of bulk fields
which have a stress-energy tensor with conformal weight $s=-4$.

Although the system of dynamical equations is now partly decoupled,
it still remains difficult to solve, for $\Omega$ depends
non-linearly on $A$, $B$ and $R$. Furthermore, it is affected by
$\mathcal{T}_{{\rm B}\mu}^\nu$ and $\mathcal{T}_{{\rm b}\mu}^\nu$.
So, let us assume that $A=A(t,r)$, $B=B(t,r)$, $R=R(t,r)$ and
$\Omega=\Omega(z)$. Then (\ref{r5DEeq}) and (\ref{5DEeqwf}) lead to
\cite{RC4} \beq {G_a^b}={\kappa_5^2}{T_{{\rm B}a}^b},\label{4DECeq}
\eeq \beq {G_5^5}={\kappa_5^2}{T_{{\rm B}5}^5},\label{5DEeqz} \eeq
\beq
6{\Omega^{-2}}{{({\partial_z}\Omega)}^2}+{\kappa_5^2}{\Omega^2}{\Lambda_{\rm
    B}}={\kappa_5^2}{U_{{\rm B}5}^5},\label{rswf1}
\eeq
\[
\left(3{\Omega^{-1}}{\partial_z^2}\Omega+{\kappa_5^2}{\Omega^2}{\Lambda_{\rm
      B}}\right){\delta_a^b}=
-{\kappa_5^2}
{\Omega^2}\delta
\left(z-{z_0}\right)\left(\lambda-{{\tilde{\mathcal{K}}}\over{\kappa_5^2}}
-f(\tilde{R}_4)\right){\delta_a^b}
\]
\beq
-{{2{\kappa_5^2}}\over{\sqrt{\tilde{g}_{55}}}}
{\Omega^2}\delta
\left(z-{z_0}\right)\left({{\tilde{\mathcal{K}}_a^b}\over{\kappa_5^2}}
+f'(\tilde{R}_4){\tilde{R}_{4a}^b}-{{\tilde{\mathcal{T}}_{{\rm
          b}a}^b}\over{2}}\right)+{\kappa_5^2}{U_{{\rm B}a}^b},\label{rswf2}
\eeq
where the latin indices represent the 4D coordinates $t$, $r$,
$\theta$ and $\phi$. Since, according to (\ref{4DECeq}) and
(\ref{5DEeqz}), $T_{{\rm B}\mu}^\nu$ depends only on $t$ and $r$,
(\ref{5DceqT}) becomes
\beq
{\nabla_a}{T_{{\rm
B}b}^a}=0.\label{5DceqT1}
\eeq
On the other hand (\ref{rswf1}) and
(\ref{rswf2}) imply that $U_{{\rm B}\mu}^\nu$ must be diagonal,
${U_{{\rm B}\mu}^\nu}=\mbox{diag}(-\bar{\rho},{\bar{p}_{\rm
r}}$\newline,${\bar{p}_{\rm
      T}},{\bar{p}_{\rm T}},{\bar{p}_5})$,
with the density $\bar{\rho}$ and pressures $\bar{p}_{\rm r}$,
$\bar{p}_{\rm T}$ satisfying $\bar{\rho}=-{\bar{p}_{\rm
r}}=-{\bar{p}_{\rm T}}$. In addition, $U_{{\rm B}\mu}^\nu$ must only
depend on $z$. Consequently, ${\nabla_a}{U_{{\rm B}b}^a}=0$ is an
identity. Then, using (\ref{u5Dceq}) and noting that ${T_{{\rm
B}\mu}^\nu}={T_{{\rm B}\mu}^\nu}(t,r)$, we find
\beq
{\partial_z}{U_{{\rm
      B}5}^5}+{\Omega^{-1}}{\partial_z}\Omega\left(2{U_{{\rm
        B}5}^5}-{U_{{\rm B}a}^a}\right)=0,\quad
2{T_{{\rm B}5}^5}={T_{{\rm B}a}^a}.\label{u5Dceq1}
\eeq
If ${U_{{\rm B}\mu}^\nu}(z)$ is a conserved tensor field like $T_{{\rm
    B}\mu}^\nu$ then
$U_{{\rm B}5}^5$ must be constant. So (\ref{u5Dceq1}) leads to the
following equations of state
\beq
2{T_{{\rm B}5}^5}={T_{{\rm
B}a}^a},\quad 2{U_{{\rm B}5}^5}={U_{{\rm B}a}^a}.\label{eqst2}
\eeq
We obtain that ${\bar{p}_5}=-2\bar{\rho}$ and $U_{B\mu}^\nu$ is
constant. On the other hand, if ${T_{{\rm B}\mu}^\nu}=\mbox{diag }
\left(-\rho,{p_{\rm r}},{p_{\rm T}},{p_{\rm T}},{p_5}\right)$ where
$\rho$, $p_{\rm r}$, $p_{\rm T}$ and $p_5$ are, respectively, the
density and pressures, then its equation of state is re-written as
\beq
\rho-{p_{\rm r}}-2{p_{\rm T}}+2{p_5}=0.\label{eqst3}
\eeq

Note that $\rho$, $p_{\rm r}$, $p_{\rm T}$ and $p_5$ must be
independent of $z$ but may be functions of $t$ and $r$. The bulk
matter is, however, inhomogeneously distributed along the fifth
dimension because the physical energy density,
${\tilde\rho}(t,r,z)$, and pressures, $\tilde{p}(t,r,z)$, are
related to $\rho(t,r)$ and $p(t,r)$ by the scale factor
$\Omega^{-2}(z)$. The bulk sector $T_{{\rm B}\mu}^\nu$ determines
the dynamics on the branes, and $U_{{\rm B}\mu}^\nu$ influences how
the gravitational field is warped around the branes. In the RS1
model $U_{{\rm B}\mu}^\nu$ also acts as a stabilizing sector
\cite{RC4}.

\section{Exact 5D warped solutions and modified gravity}

The $\rm{AdS}_5$ braneworld dynamics are defined by the solutions of
(\ref{4DECeq}) to (\ref{5DceqT1}) and (\ref{eqst3}). Let us first
solve (\ref{rswf1}) and (\ref{rswf2}). As we have seen $U_{{\rm
B}\mu}^\nu$ is constant with $\bar{\rho}=-{\bar{p}_{\rm
r}}=-{\bar{p}_{\rm T}}=-{\bar{p}_5}/2$. After integrating
(\ref{rswf1}), by using the cartesian coordinate $y$ related to $z$
by $z=l{e^{y/l}}$, and taking into account the $Z_2$ symmetry, we
obtain \cite{RC4} \beq \Omega(y)={e^{-|y|/l}}\left(1+{p_{\rm
B}^5}{e^{2|y|/l}}\right),\label{wfp5y} \eeq where $l$ is the AdS
radius given by $l=1/\sqrt{-{\Lambda_{\rm B}}{\kappa_5^2}/6}$ and
${p_{\rm B}^5}={\bar{p}_5}/(4{\Lambda_{\rm B}})$. This set of
solutions must also satisfy (\ref{rswf2}) which contains the Israel
jump conditions. These boundary conditions imply that the brane
energy-mo\-mentum $\tilde{\mathcal{T}}_{{\rm b}a}^b$ is given by
\beq {\tilde{\mathcal{T}}_{{\rm b}a}^b}=\left[{\lambda_{\rm
RS}}\left(R-2\right)+R\left(\lambda-{{\tilde{\mathcal{K}}}
\over{\kappa_5^2}}-f(\tilde{R}_4)\right)\right]{\delta_a^b}
+2\left({{\tilde{\mathcal{K}}_a^b}\over{\kappa_5^2}}
+f'(\tilde{R}_4){\tilde{R}_{4a}^b}\right),\label{4Dbeq} \eeq where
${\lambda_{\rm RS}}=6/(l{\kappa_5^2})$ and $R=\Omega(0)={p_{\rm
        B}^5}+1$. ${\tilde{\mathcal{T}}_{{\rm b}a}^b}$,
${\tilde{\mathcal{K}}_a^b}$ and ${\tilde{R}_{4a}^b}$ are calculated at
the brane boundary $y=0$.

To determine the dynamics on the brane we need to solve
(\ref{4DECeq}) and (\ref{5DEeqz}) when $T_{{\rm B}\mu}^\nu$
satisfies (\ref{5DceqT1}) and (\ref{eqst3}). First note that if
$T_{{\rm B}\mu}^\nu=0$ then the 5D metric is given by \beq
d{\tilde{s}_5^2}=d{y^2}+{\Omega^2}d{\Omega_4^2}, \eeq where
$d{\Omega_4^2}=d{\chi^2}+{\sin^2}\chi d{\Omega_3^2}$ with
$d{\Omega_3^2}=d{\xi^2}+{\sin^2}\xi d{\Omega_2^2}$. The boundary
conditions (\ref{4Dbeq}) then imply that \beq {\tilde{T}_{{\rm
b}a}^b}=\left[{{\lambda_{\rm
RS}}\over{3R}}(R-2)(R+1)+R\left(\lambda-f\left({{12}\over{R^2}}\right)\right)
+{6\over{R^2}}f'\left({{12}\over{R^2}}\right)\right]{\delta_a^b}.
\eeq Since the boundary condition depends on the arbitrary function
$f$, the matter confined to the brane is ambiguous. To lift such an
ambiguity some fundamental physical principle should be added to the
RS scenario, in order to fix the corresponding theory of gravity on
the brane \cite{NO}.

With a nonzero $T_{{\rm  B}\mu}^\nu$ it is possible to consider more
general 5D geometries. Let us first note that as long as $p_5$
balances $\rho, {p_{\rm  r}}$ and $p_{\rm T}$ according to
(\ref{5DEeqz}) and  (\ref{eqst3}), the 4D equation of state of the
bulk fluid is not constrained. In previous work, three examples
corresponding to inhomogeneous dust, generalized dark radiation and
homogeneous polytropic matter were analyzed \cite{RC2,RC3,RC4}. The
latter describes the dynamics on the brane of dark energy in the
form of a polytropic fluid. The diagonal conformal matter can be
defined by \beq \rho={\rho_{\rm P}}+\Lambda,\; {p_{\rm
r}}+\eta{{\rho_{\rm P}}^\alpha}+\Lambda=0,\;{p_{\rm T}}={p_{\rm
r}},\;{p_5}=-{1\over{2}} \left({\rho_{\rm P}} +3\eta{{\rho_{\rm
P}}^\alpha}\right)-2\Lambda, \label{dmeqst} \eeq where $\rho_{\rm
P}$ is the polytropic energy density, $\Lambda$ is a bulk quantity
which mimics a brane cosmological constant and the parameters
($\alpha$, $\eta$) characterize different polytropic phases.

Solving the conservation equations, we find \cite{RC3,BBS} \beq
{\rho_{\rm P}}={{\left(\eta+{a\over{S^{3-3\alpha}}}\right)}^
{1\over{1-\alpha}}},\label{dmdena} \eeq where $\alpha\not=1$, $a$ is
an integration constant and $S=S(t)$ is the Robertson-Walker scale
factor of the braneworld which is related to the physical radius by
$R=rS$. For $-1\leq\alpha<0$ the fluid is in its generalized
Chaplygin phase (see also \cite{BBS}). With this density the
Einstein equations lead to the following 5D dark energy polytropic
solutions \cite{RC3} \beq
d{\tilde{s}_5^2}={\Omega^2}\left[-d{t^2}+{S^2}
\left({{d{r^2}}\over{1-k{r^2}}}+{r^2}d{\Omega_2^2}\right)\right]+d{y^2},
\label{dmsol1} \eeq where the brane scale factor $S$ satisfies \beq
{\ddot{S}\over{S}}=-{{\kappa_5^2}\over{6}}\left({\rho_{\mbox{\tiny
P}}}-3\eta{{\rho_{\mbox{\tiny
          P}}}^\alpha}-2\Lambda\right)\label{dmdyeq}
\eeq
and
\beq
{\dot{S}^2}={{\kappa_5^2}\over{3}}\left({\rho_{\mbox{\tiny
        P}}}+\Lambda\right){S^2}-k.\label{dmdyeqa}
\eeq
The global evolution of the observable
universe is then given by \cite{RC1,RC3}
\beq
S{\dot{S}^2}=V(S)={{\kappa_5^2}\over{3}}\left[{{\left(\eta{S^{3-3\alpha}}+
a\right)}^{1\over{1-\alpha}}}+\Lambda{S^3}\right]-k S.\label{dmdineq}
\eeq

Let us now consider the boundary conditions (\ref{4Dbeq}) which
define the brane tensor $\tilde{\mathcal{T}}_{{\rm b }a}^b$. Using
the metric (\ref{dmsol1}) we obtain
\beq
{\tilde{\mathcal{T}}_{{\rm
b}t}^t}={{\lambda_{\rm
      RS}}\over{3R}}\left(R-2\right)\left(R+1\right)
      +R\left[\lambda-f(\tilde{R}_4)\right]+{6\over{R^2}}
{{\ddot{S}}\over{S}}f'(\tilde{R}_4),\label{4Dbe1}
\eeq
\beq
{\tilde{\mathcal{T}}_{{\rm b}r}^r}={{\lambda_{\rm
      RS}}\over{3R}}\left(R-2\right)\left(R+1\right)
      +R\left[\lambda-f(\tilde{R}_4)\right]+{2\over{R^2}}f'(\tilde{R}_4)
\left({{\ddot{S}}\over{S}}+2{{{\dot{S}^2}+k}\over{S^2}}\right),\label{4Dbe2}
\eeq
where $\tilde{R}_4=6(S\ddot{S}+\dot{S}^2+k)/({S^2}R)$. We also
have ${\tilde{\mathcal{T}}_{{\rm b}\theta}^\theta}
={\tilde{\mathcal{T}}_{{\rm b}\phi}^\phi}={\tilde{\mathcal{T}}_{{\rm
b}r}^r}$. All the other components of ${\tilde{\mathcal{T}}_{{\rm
b}\mu}^\nu}$ are equal to zero. Subtracting (\ref{4Dbe1}) and
(\ref{4Dbe2}) we find
\beq
{\tilde{\mathcal{T}}_{{\rm
b}t}^t}-{\tilde{\mathcal{T}}_{{\rm
b}r}^r}={4\over{R^2}}f'(\tilde{R}_4)\left({{\ddot{S}}\over{S}}
-{{{\dot{S}^2}+k}\over{S^2}}\right).\label{4Dbe3}
\eeq
If $f=0$ the
theory of gravity on the brane is that of Einstein. Then we obtain
${\tilde{\mathcal{T}}_{{\rm
      b}t}^t}={\tilde{\mathcal{T}}_{{\rm b}r}^r}$ which implies that
\beq
{\tilde{\mathcal{T}}_{{\rm b}a}^b}=\left[{{\lambda_{\rm
      RS}}\over{3R}}\left(R-2\right)\left(R+1\right)+R\lambda\right]{\delta_a^b}.
\eeq
For ${\tilde{\mathcal{T}}_{{\rm b}\mu}^\nu}=0$ we obtain the brane
tension $\lambda={\lambda_{\rm
      RS}}(2-R)(R+1)/(3{R^2})$. Note that if $f$ is a constant then
  the same is true with the appropriate shift in the brane tension,
  $\lambda\to\lambda-f$. If $f$ is not constant and there is no matter confined to the brane
or if ${\tilde{T}_{{\rm b}t}^t}={\tilde{T}_{{\rm b}r}^r}$ then we are
lead to
\beq
{{\ddot{S}}\over{S}}={{{\dot{S}^2}+k}\over{S^2}}.\label{eq4}
\eeq
Using (\ref{dmdyeq}) and (\ref{dmdyeqa}) we find that equation (\ref{eq4})
restricts ${\rho_{\rm P}}$ to be zero or a constant, ${\rho_{\rm
    P}}={\eta^{1/(1-\alpha)}}$. Under this circumstances, the potential $V(S)$ is given by
\beq
V(S)={{\kappa_5^2}\over{3}}\left({\rho_{\mbox{\tiny
        P}}}+\Lambda\right){S^3}-kS,
\eeq
and describes a global evolution determined by an effective
brane cosmological constant $\Lambda$ or $\Lambda+{\eta^{1/(1-\alpha)}}$.

In general, when $f$ is not constant and ${\tilde{\mathcal{T}}_{{\rm b}t}^t}$ is different from
${\tilde{\mathcal{T}}_{{\rm b}r}^r}$, the boundary equations
(\ref{4Dbe1}) and (\ref{4Dbe2}) define state constraints relating
the theory of gravity represented by $f$ and the matter confined to
the brane characterized by ${\tilde{\mathcal{T}}_{{\rm
      b}\mu}^\nu}$. For example if $f$ is given by (\ref{spf}) we find
\beq
{\tilde{\mathcal{T}}_{{\rm
b}t}^t}={{\lambda_{\rm
      RS}}\over{3R}}\left(R-2\right)\left(R+1\right)
      +R\left(\lambda+{a\over{\tilde{R}_4}}-b{\tilde{R}_4^2}\right)+{6\over{R^2}}
{{\ddot{S}}\over{S}}\left({a\over{\tilde{R}_4^2}}+2b{\tilde{R}_4}\right),\label{4Dbe4}
\eeq
\beq
{\tilde{\mathcal{T}}_{{\rm
b}t}^t}-{\tilde{\mathcal{T}}_{{\rm
b}r}^r}={4\over{R^2}}\left({a\over{\tilde{R}_4^2}}+2b{\tilde{R}_4}\right)\left({{\ddot{S}}\over{S}}
-{{{\dot{S}^2}+k}\over{S^2}}\right).\label{4Dbe5}
\eeq
In this perspective, the gravitational brane dynamics are defined by
the conformal bulk fields of weight -4 and are supported on the
brane by a simple brane tension, if $f$ is a constant,
and by more general matter, if $f$ is some
arbitrary function of $\tilde{R}_4$. Such dynamics may be of the
dark energy polytropic type or even more general and are independent
of $f$. What actually depends on $f$ is the kind of matter that is
confined to the brane. As is clear, in (\ref{4Dbeq}) each $f$ is
balanced by a certain ${\tilde{\mathcal{T}}_{{\rm b}\mu}^\nu}$ which
in general must be associated with non-conformal fields.

\section{Conclusions}

We have analyzed in this paper exact 5D solutions which describe the
dynamics of $\mbox{AdS}_5$ braneworlds when conformal fields of
weight -4 propagate in the bulk. We have considered the possibility
of a general deviation of the theory of gravity on the brane from
the Einstein form and analyzed the effect this can have on the
capability of the $\mbox{AdS}_5$ braneworld to describe on the brane
the dynamics of inhomogeneous dust, generalized dark radiation and
homogeneous polytropic dark energy, respectively. These solutions
are determined by the conformal bulk fields of weight -4. While a
sector of the fields generates the dynamics on the brane, another
sector affects the way how gravity is warped around the brane. In
the RS1 model, this latter sector also provides a way to stabilize
the radion mode. In this work we have shown that, within our model
of modified gravity, the existence of such exact 5D geometries
requires the propagation of non-conformal matter fields confined to
the brane. The corresponding energy-momentum tensor is defined in
terms of the gravitational elements composing the relevant theory of
gravity on the brane. Thus, these non-conformal fields play the role
of supporting the dynamics on the brane when modified gravity is
present. Their state explicitly depends on how gravity deviates from
the Einstein form and on the $\mbox{AdS}_5$ braneworld dynamics
generated by the conformal bulk fields. In this context, the
ambiguity associated with the brane theory of gravity is reflected
on the kind of matter that is confined to the brane. Note also that,
related with this, one can  consider a generalized non-minimal
coupling of gravity with matter on the brane, of the form
$f(\tilde{R}_4){\tilde{\mathcal{L}}_{\rm d}}$, where
$\tilde{\mathcal{L}}_{\rm d}$ is some matter lagrangian including
also a kinetic term (see \cite{NOA}). In the case of usual 4D
gravity, this provides an explanation for dark energy dominance.
Such effect on the brane will be studied elsewhere. \vspace{0.5cm}

\noindent{\bf Acknowledgements} \vspace{0.25cm}

We would like to thank Sergei Odintsov and Cenalo Vaz for helpful
discussions. This paper has been supported by
{\it Centro de Electr\'onica, Optoelectr\'onica e Telecomunica\c
{c}\~oes} (CEOT), by {\it Funda\c {c}\~ao
para a Ci\^encia e a Tecnologia} (FCT) and {\it Fundo Social
Europeu} (FSE), contract SFRH/BPD\-/7182/2001 ({\it III Quadro
Comunit\'ario de Apoio}), by DGICYT (Spain), project BFM2003-00620,
and by {\it Conselho de Reitores das Universida}\-{\it des Portuguesas}
(CRUP) and {\it Ministerio de Educaci\'on y Ciencia} (MEC, Spain),
projects E-126/04 and HP2003-0145.


\begin{thebibliography}{30}

\bibitem{RS1}
L. Randall and R. Sundrum, Phys. Rev. Lett. {\bf 83}, 3370 (1999)

\bibitem{RS2}
L. Randall and R. Sundrum, Phys. Rev. Lett. {\bf 83}, 4690 (1999)

\bibitem{GW}
W. D. Goldberger and M. B. Wise, Phys. Rev. D. {\bf 60}, 107505
(1999)

W. D. Goldberger and M. B. Wise, Phys. Rev. Lett. {\bf 83}, 4922
(1999)

W. D. Goldberger and M. B. Wise, Phys. Lett. B {\bf 475}, 275 (2000)

\bibitem{NK}
N. Kaloper, Phys. Rev. D {\bf 60}, 123506 (1999)

T. Nihei, Phys. Lett. B {\bf 465}, 81 (1999) 

C. Cs\'aki, M. Graesser, C. Kolda and J. Terning, Phys. Lett. B {\bf
  462}, 34 (1999) 

J. M. Cline, C. Grojean and G. Servant, Phys. Rev. Lett. {\bf 83}, 4245 (1999)

\bibitem{KKOP}
P. Kanti, I. I. Kogan, K. A. Olive and M. Pospelov, Phys. Lett. B {\bf
  468}, 31 (1999) 

P. Kanti, I. I. Kogan, K. A. Olive and M. Pospelov, Phys. Rev. D {\bf 61}, 106004 (2000)

\bibitem{WFGK}
O. DeWolf, D. Z. Freedman, S. S. Gubser and A. Karch, Phys. Rev. D
{\bf 62}, 046008 (2000)

\bibitem{GT}
J. Garriga and T. Tanaka, Phys. Rev. Lett. {\bf 84}, 2778 (2000)

\bibitem{GKR}
S. Giddings, E. Katz and L. Randall, J. High Energy Phys. {\bf 03}, 023 (2000)

\bibitem{CGRT}
C. Cs\'aki, M. Graesser, L. Randall and J. Terning, Phys. Rev. D {\bf
  62}, 045015 (2000)

\bibitem{SNSO}
S. Nojiri and S. D. Odintsov, Phys. Lett. B {\bf 484},
119 (2000)

\bibitem{TM}
T. Tanaka and X. Montes, Nucl. Phys. {\bf B582}, 259 (2000)

\bibitem{BC}
B. Carter, Phys. Rev. D {\bf 48}, 4835 (1993)

R. Capovilla and J. Guven, Phys. Rev. D {\bf 51}, 6736 (1995)

R. Capovilla and J. Guven, Phys. Rev. D {\bf 52}, 1072 (1995)

\bibitem{SMS}
T. Shiromizu, K. I. Maeda and M. Sasaki, Phys. Rev. D {\bf 62}, 024012
(2000)

M. Sasaki, T. Shiromizu and K. I. Maeda, Phys. Rev. D {\bf 62}, 024008 (2000)

\bibitem{COSp}
J. Garriga and M. Sasaki, Phys. Rev. D {\bf 62}, 043523 (2000)

R. Maartens, D. Wands, B. A. Bassett and I. P. C. Heard, Phys. Rev.
D {\bf 62}, 041301 (2000)

H. Kodama, A. Ishibashi and O. Seto, Phys. Rev. D {62}, 064022 (2000) 

D. Langlois, Phys. Rev. D {\bf 62}, 126012 (2000) 

C. van de Bruck, M. Dorca, R. H. Brandenberger and A. Lukas,
Phys. Rev. D {\bf 62}, 123515 (2000)

K. Koyama and J. Soda, Phys. Rev. D {\bf 62}, 123502 (2000)

\bibitem{DMPR}
N. Dadhich, R. Maartens, P. Papadopoulos and V. Rezania, Phys.
Lett. B {\bf 487}, 1 (2000)

N. Dadhich and S. G. Ghosh, Phys. Lett. B {\bf 518}, 1 (2001) 

C. Germani and R. Maartens, Phys. Rev. D {\bf 64}, 124010 (2001)

M. Bruni, C. Germani and R. Maartens, Phys. Rev. Lett. {\bf 87}, 231302 (2001)

\bibitem{RM}
R. Maartens, Phys. Rev. D {\bf 62}, 084023 (2000)

\bibitem{RC1}
R. Neves and C. Vaz, Phys. Rev. D {\bf 66}, 124002 (2002)

\bibitem{RC2}
R. Neves and C. Vaz, Phys. Rev. D {\bf 68}, 024007 (2003)

\bibitem{RC3}
R. Neves and C. Vaz, Phys. Lett. B {\bf 568}, 153 (2003)

\bibitem{RN}
R. Neves, TSPU Vestnik Natural and Exact Sciences 7, 94 (2004)

\bibitem{RC4}
R. Neves and C. Vaz, J. Phys. A: Math. Gen. 39, 6617 (2006)

\bibitem{EONO}
E. Elizalde, S. Nojiri, S.D. Odintsov and S. Ogushi, Phys. Rev. D
{\bf 67}, 063515 (2003)

S. Nojiri and S. D. Odintsov, JCAP {\bf 06}, 004 (2003) 

E. Elizalde, S. Nojiri and S.D. Odintsov, Phys. Rev. D {\bf 70}, 043539 (2004)

\bibitem{NO1}
S. Nojiri and S. D. Odintsov, Phys. Rev. D {\bf 68}, 123512 (2003)

\bibitem{modg}
M. C. B. Abdalla, S. Nojiri and S. D. Odintsov, Class. Quantum
Grav. {\bf 22}, L35 (2005)

\bibitem{NO}
S. Nojiri and S. D. Odintsov, Gen. Rel. Grav. {\bf 37}, 1419 (2005)

\bibitem{BBS}
M. C. Bento, O. Bertolami and S. S. Sen, Phys. Rev. D {\bf 66}, 043507 (2002)

M. C. Bento, O. Bertolami and S. S. Sen, Phys. Rev. D {\bf 67}, 063003
(2003)

M. C. Bento, O. Bertolami and S. S. Sen, Phys. Rev. D {\bf 70}, 083519
(2004)

M. C. Bento, O. Bertolami, N. M. C. Santos and S. S. Sen, Phys. Rev. D {\bf 71}, 063501 (2005)

\bibitem{NOA}
S. Nojiri and S. D. Odintsov, Phys. Lett. B {\bf 599}, 137 (2004)

G. Allemandi, A. Borowiec, M. Francaviglia and S. D. Odintsov,
Phys. Rev. D {\bf 72}, 063505 (2005)

\end{thebibliography}
\end{document}